\documentclass[
 twocolumn,floatfix,showpacs,preprintnumbers,
 amsmath,amssymb,pre]{revtex4}

\usepackage{graphicx}
\usepackage{amsmath}
\usepackage{hyperref}
\usepackage{wasysym}
\usepackage{bm}
\usepackage{isolatin1}

\renewcommand{\v}[1]{\boldsymbol{#1}}

\newcommand{\PROM}[1]{\left\langle #1\right\rangle}

\DeclareGraphicsExtensions{.eps}

\begin{document}

\setlength{\voffset}{3\baselineskip}

\title{Synchronization universality classes and stability of smooth coupled map 
 lattices}

\author{Franco Bagnoli$^{1,2,3}$, Ra\'ul Rechtman$^4$\\
{\small\it
 $^1$Dipartimento di Energetica, Universit\`a di Firenze,\\
 Via S. Marta 3, I--50139 Firenze, Italy,
 \email{franco.bagnoli@unifi.it} \\
 $^2$Centro Interdipartimentale per lo Studio delle Dinamiche
 Complesse, Firenze,\\
 $^3$INFN Firenze\\
 $^4$Centro de Investigaci\'on en Energ\'\i a, Universidad Nacional
 Aut\'onoma\\
 de M\'exico, Apdo. Postal 34, 62580 Temixco, Mor., Mexico\\
 }}

\date{\today}

\begin{abstract}
 We study two problems related to spatially extended systems: the
 dynamical stability and the universality classes of the replica
 synchronization transition. We use a  simple model of one dimensional
 coupled map lattices and show that chaotic behavior implies that the
 synchronization transition belongs to the multiplicative noise
 universality class, while stable chaos implies that the
 synchronization transition belongs to the directed percolation
 universality class.  
\end{abstract}


\maketitle

\section{Introduction}

Chaos is a widely studied phenomenon. It represents the generic
behavior of nonlinear systems and it is also assumed to be \emph{the}
microscopic mechanism for the amplification of uncertainties about
initial conditions. The exponential growth of an initial small
uncertainty is measured by the maximum Lyapunov exponent (MLE).
However, in high-dimensional systems, and especially in extended
systems with short-range coupling, the origin of uncertainty may be
different from the previous one.  To be more specific, the systems we
take into considerations are defined on a lattice with short range
interactions.

Just to give a simple example of the different role of chaos in
low-dimensional with  respect to extended systems, let us consider an
array of \emph{uncoupled} chaotic  maps. This is clearly a chaotic
system; a localized uncertainty will grow in time and will remain
localized. If there is a coupling between neighboring maps, a
localized uncertainty will propagate to neighboring sites. This
propagation is a  linear phenomenon, and is hidden by the exponential
amplification due to the chaotic  dynamics of maps. If we now consider
\emph{non-chaotic} (\emph{i.e.}, stable) coupled maps, only the
spreading mechanism may be present. Since the individual maps  are
stable, infinitesimal uncertainties are always absorbed, but
\emph{finite} ones  may survive. One of the most striking examples are
\emph{class-3} cellular automata  (CA)~\cite{wolfram86}, that are
equivalent to superstable maps, but may nevertheless  be denoted
``chaotic'' because a localized uncertainty may spread to the whole
system.

While cellular automata may be seen as coupled maps of extreme type,
that only  take values zero and one, CA-like behavior  has been
observed also in  continuous maps, albeit with discontinuities or
extremely sharp transitions~\cite{politi93}. In these cases the
maximum Lyapunov exponents (if defined) is negative but defects and
information may propagate and amplify in a non-exponential way. When
finite, these systems are generally periodic, but the period grows at
least exponentially with the size, so that the transient is  the only
observable state~\cite{stable,politi93,cecconi98}. This behavior is
denoted as \emph{stable chaos}. 

As mentioned before, evaluation of the MLE is the most direct method
to establish the stability properties of a system but is somewhat
limited for extended systems as  coupled map lattices (CMLs). An
alternative method for the measurement of  uncertainty is based on the
synchronization of replicas of a  system~\cite{Pecora,Pikovski}.
Typically, one fraction of each replica is added to the other and the
distance between them vanishes as this fraction 
varies~\cite{ahlers02,DrozLipowski}. Another mechanism involves the
addition of the  same external noise to both replicas~\cite{baroni}.
In both cases, there is a control parameter $p$ that measures the
\emph{strength} of the synchronization.

We may have different scenarios, according with the degree of 
unpredictability of the system. Chaotic systems are expected to
amplify the distance between replicas. For a value of $p$ slightly
below the synchronization threshold, some patches may synchronize for
some time, after which they will  separate. This picture resembles
that of a growing interface that may stay pinned to local traps. From
field theory studies,  such a behavior is denoted \emph{multiplicative
noise} (MN) and is equivalent   to the behavior of the (bounded)
Kardar-Parisi-Zhang equation, which describes the  behavior of a
growing surface that tends to pin and is pushed from 
below~\cite{munoz,mult-noise,kardar,ginelliDepinning,LipowskiDroz}. On the
other hand, stable systems have a negative MLE. So, replicas should
naturally synchronize once their distance is (locally) below the
threshold of validity of linear analysis. However, when the (local)
difference is large, non-linear terms may maintain or amplify this
distance. In this case synchronized patches may be destabilized only
at the boundaries. Again, theoretical studies associate  such a
behavior to that of directed percolation 
(DP)~\cite{ahlers02,grassberger99,ginelliMNDP}. In other words, one
may illustrate this behavior by saying that the synchronized state is
similar to the \emph{void} state of the field theory descriptions, and
that this void state is absorbing in the second case and unstable in
the first case~\cite{munozpastor}. 

It has been noted that CMLs with continuous chaotic maps lead to
synchronization with a MN character while the synchronization of CMLs
where the local map is  discontinuous~\cite{Grassberger,ahlers02}, or
with very sharp transitions~\cite{DrozLipowski}, belong to the DP
universality class. The use of simple models where the transition
between the two behaviors may be triggered  by a variation of control
parameter has been addressed for the  stochastic synchronization of
CMLs in Ref.~\cite{ginelliMNDP} and in growth  models in
Refs.~\cite{ginelliDepinning,LipowskiDroz}. Randomness can be safely
used if the system is chaotic although the systems under study are
deterministic.  This is not the case when there is stable chaos,
unless coarse graining is used.

In what follows we investigate the relation between stability  and 
the synchronization character of CMLs. We present a model that, by
changing  continuously a single parameter, can vary smoothly from
\emph{chaotic}  to  \emph{chaotically stable} (non chaotic but
unpredictable) behavior,   and finally to \emph{cellular automata}
behavior.  From the point of view of the synchronization transition,
the model exhibits a  transition from the MN universality class to
that of DP. The transition from  chaos to stable chaos occurs at the
same value of the parameter as the transition from the MN to the DP
universality class. Moreover, we show that these transitions  occur
for smooth and continuous maps, and that at the transition  stable 
synchronized trajectories are still present.  The transition from
standard chaos to stable chaos occurs through  the appearance of
transient chaos, {\emph i.e.}, the system in the stable chaotic phase
exhibits a transient chaotic behavior before being attracted into the
stable-chaos trajectory.

The paper is organized as follows. In Sec.~\ref{sec:model} we present
the CML model and the synchronization mechanism. In the following section we present
the main numerical results and in Sec.~\ref{sec:conclusions} we end with 
some conclusions and directions for further work.

\section{The model}
\label{sec:model}

A coupled map lattice $F$ defines a flow on $[0,1]^N$, 
$\v{x}^{t+1}=F(\v{x}^t)$ with $\v{x}=(x_0,\dots,x_{N-1})$ and 
$t=0,1,\dots$. It is defined locally by  \begin{equation}
x_i^{t+1}=f((1-2\epsilon)x_i^t+\epsilon(x_{i-1}^t+x_{i+1}^t))	
\end{equation} with $i=0,\dots,N-1$ and periodic boundary conditions, 
$0\leq\epsilon\leq 1/2$ and $f:[0,1]\to[0,1]$. In what follows
$\epsilon=1/3$ so that  
\begin{equation}
 \label{eq:cml}
 x_i^{t+1}=f(\sigma_i^t)
\end{equation}
with $\sigma_i=(x_{i-1}+x_i+x_{i+1})/3$. 

The model is defined by the choice of $f$ that depends  on a parameter
$a$ such that for large values of this parameter, the CML behaves as
an elementary cellular automaton (super-stable system),  and in the
other limit the individual map is chaotic. One way to define $f$  is
by 
\begin{equation}
 \label{f}
 f(x;a)=
 \begin{cases}
  (6x)^a/2		& 0\leq x < 1/6,\\
  1-|6(1/3-x)|^a/2	& 1/3\leq x < 1/2,\\
  |6(x-2/3)|^a/2	& 1/2\leq x < 5/6,\\
  1-(6(1-x))^a/2	& 5/6\leq x < 1,
 \end{cases}        
\end{equation}
where $1\le a<\infty$. This function was introduced in
Ref.~\cite{stablesync} in  the context of stochastic synchronization. 

\begin{figure}
 \includegraphics[width=\columnwidth]{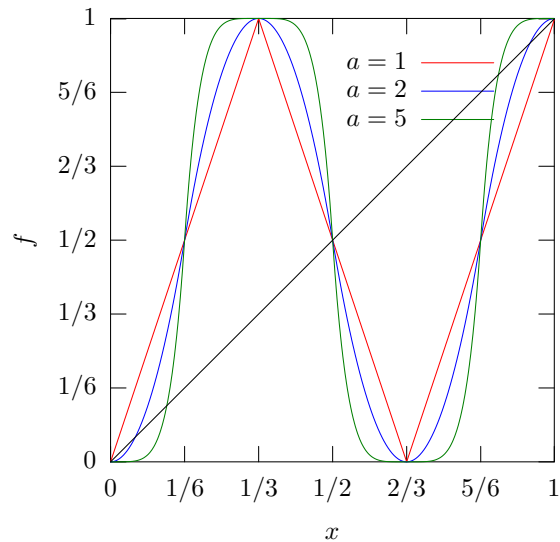}
 \caption{\label{fig:fa_10} (Color Online) The graph of $f_(x)$ for three
values 
  of $a$.} 
\end{figure}
\begin{figure}
 \includegraphics[width=\columnwidth, height=\columnwidth]{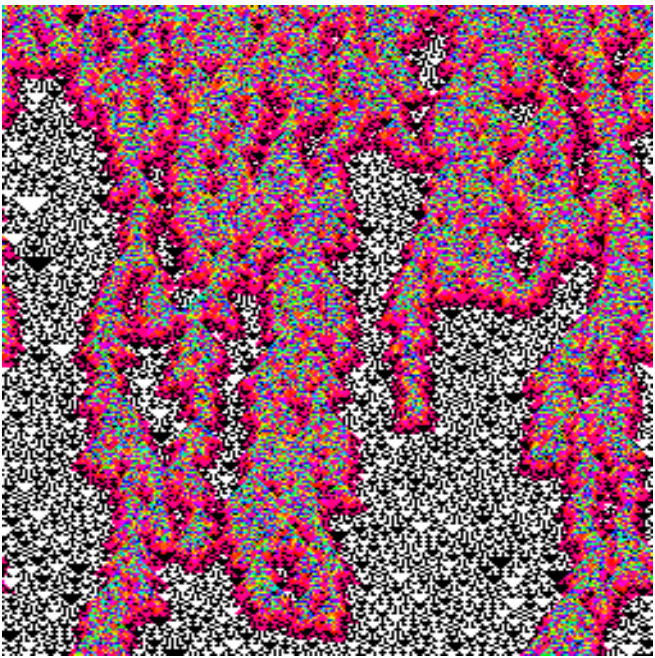}
 \vspace{.2cm}
 \caption{\label{fig:cml} (Color Online) Space time pattern of the CML
  of  Eq.~(\protect{\ref{eq:cml}}) with $a=1.9$ and $N=256$ drawn
  horizontally for a total of $T=300$ time steps drawn vertically from
  top to bottom. The initial configuration $\v{x}^0$ is chosen
  randomly. The color code assigns white (black) whenever
  $x_i^t=0\,(1)$ and a rainbow color scale for other values of $x_i^t$
  starting with red for values near zero. Patches of CA behavior 
  (rule 150) appear after a short transient and will eventually fill
  the whole  pattern.}
\end{figure}

The graph of $f$ is shown in Fig.~\ref{fig:fa_10} and a typical space
time pattern is shown in Fig.~\ref{fig:cml}. For a choice of the
parameter  $a\gtrsim 1.8$, the CML is asymptotically equivalent to a
cellular  automaton (CA) with triangles of zeroes and ones. The
function $f$ was chosen in  such a way that the CML in the limit of
large $a$ will behave as the ``chaotic'' elementary CA rule
150~\cite{wolfram86}. 

Chaotic systems are characterized by the divergence of initially  near
trajectories, and a quantitative indicator for chaotic motion is
(generally) a positive maximum Lyapunov exponent (MLE)
$\lambda(\boldsymbol{x}^0)$,  that depends on the trajectory. In order
to probe the configuration space, we  numerically evaluate
$\lambda_{M,T}$, the average over $M$ initial randomly  chosen
configurations $\v{x}^0$ of finite time Lyapunov exponents during $T$
time steps.

We study the synchronization of CMLs by considering the following scheme:
\begin{align}
 x_i^{t+1}&=f(\sigma_i^t),						\\
 y_i^{t+1}&=(1-p)f(\tau_i^t)+pf(\sigma_i^t),				\\
 u_i^{t+1}&=\left|x_i^{t+1}-y_i^{t+1}\right|					\\
          &=(1-p)\left|f(\sigma_i^t)-f(\tau_i^t)\right|,	
\end{align}
where $\tau_i=(y_{i-1}+y_i+y_{i+1})/3$. 
Synchronization occurs when $u_i^t\to 0$ $\forall i$, 
which is equivalent to a vanishing norm $u^t$ of $\v{u}^t$
defined by
\begin{equation}
 u^t=\dfrac{1}{N}\sum_i u_i^t.
\end{equation}
The synchronization threshold is called $p_s$.  In numerical
simulations, we usually  evaluate $\PROM{u}$, defined as an average
over $T$ time steps after a relaxation  time $t_{rel}$ which is then
averaged over $M$ random initial conditions.  In
Fig.~\ref{fig:sync-cml-00-gr}  we show $u^t$ for several values of $a$
and $p_s$.

The linear stability of $u$ is measured by the transverse Lyapunov
exponent(TLE) $\lambda_{\bot}$ defined by~\cite{PecoraCarroll} 
\begin{equation}
 \lambda_{\bot}(\v{x}^0)=\log(1-p)+\lambda(\v{x}^0).
\end{equation}
The synchronized state becomes stable with respect to infinitesimal
perturbation  when $\lambda_\bot=0$ so that the value of $p$ at
synchronization, $p_\ell$, is  given in the linear approximation by
\begin{equation}
 \label{eq:ps}
 p_\ell=1-\exp(-\lambda).
\end{equation}
This does not mean that synchronization actually takes place for this
value of $p$.  It may happen that the system stays unsynchronized for
$p>p_\ell$. In  Ref~\cite{ahlers02} this behavior has been associated
to discontinuities of the single map taking as reference the Bernoulli
shift. We shall show in the following  that even smooth, non-chaotic
maps may correspond to $p_s>p_\ell$.

\section{Results}
\label{sec:results}

We first discuss the stability properties of the CML presented in the
previous section and then the character of the synchronization
transition. If $1<a<a_c$ the CML is chaotic and if $a>a_c$ the CML
exhibits stable chaos. We found $a_c=1.8142(2)$ as discussed below.
For $1<a<a_c$ almost any trajectory starting at any initial state
$\v{x}^{0}$ with $x_i^0$ chosen randomly between zero and one  has a
positive MLE. However, if the initial state is such that $x_i^0=0,1$
randomly, the CML will evolve as the elementary CA rule 150 with a
negative MLE (actually $\lambda=-\infty$). 

The configuration space of the system is the unitary hypercube
$[0,1]^N$, whose  vertices correspond to automata configurations
(zeroes and ones). Starting from  one vertex, the time evolution will
visit only vertices. There is an attractor near  the vertices of the
hypercube that brings any trajectory with initial state in this
attractor toward these trajectories CA trajectories. The transition
from chaos  to stable-chaos corresponds to the transition where this
attractor fills the  hypercube.

\begin{figure}
 \hskip -5mm\includegraphics[width=\columnwidth]{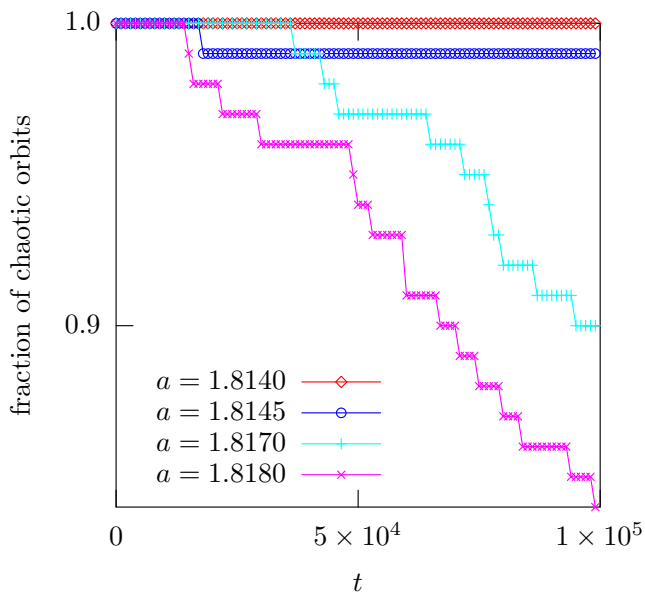}
 \caption{\label{fig:cml07-1814-1818} (Color online) Fraction of chaotic 
  orbits as a function of time for different values of $a$ for a total 
  of $M=100$ random initial states, $N=1,024$ and $T=100N$.}
\end{figure}

For $a>a_c$ the system exhibits transient chaos: initially nearby
trajectories  diverge and are finally attracted to a CA type
evolution. Numerically one observes a  positive finite-time MLE that
(suddenly) becomes negative. This fact is used to find the value of
$a_c$ as shown in  Fig.~\ref{fig:cml07-1814-1818}. For different
values of $a$, the MLE of $M$ orbits starting from randomly chosen
initial states is evaluated  and the fraction of  chaotic orbits is
shown as a function of time. The first value of $a$ for which this
fraction is not one is taken as $a_c$. For $a>a_c$ the fraction of
chaotic orbits goes to zero as $t\to\infty$ so the measure of the set
of initial states showing stable chaos goes to one. 

\begin{figure}
\begin{tabular}{cc}
 \includegraphics[width=0.5\columnwidth]{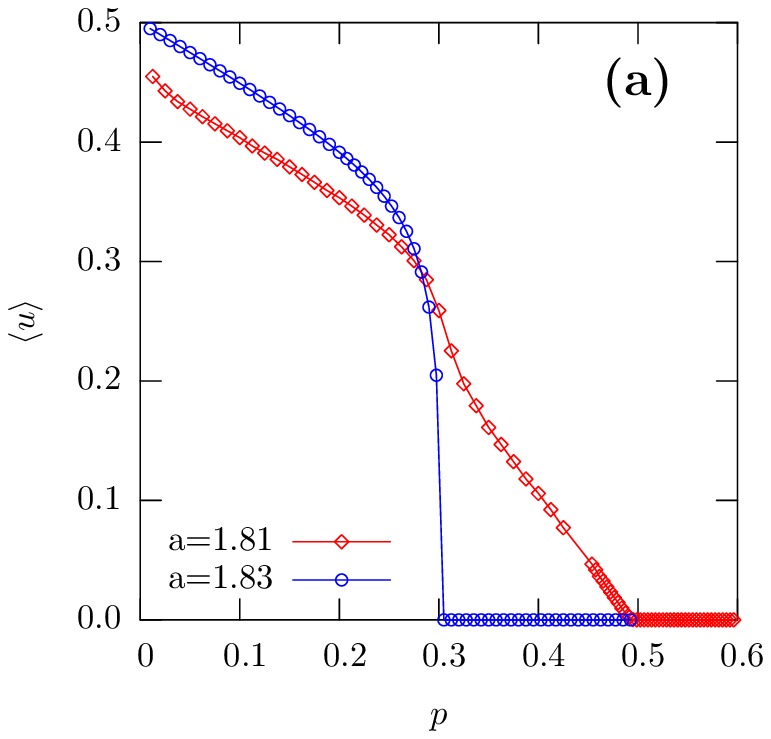}&
 \includegraphics[width=0.5\columnwidth]{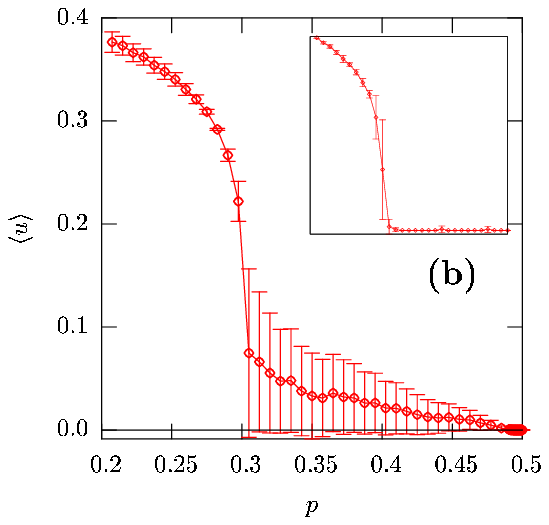}
\end{tabular}
 \caption{\label{fig:s02} (Color online) Dependence of the order parameter 
  $\PROM{u}$ on $p$ near the critical value $a_c$. In (a) 
  $T=t_{rel}=5\times 10^4$, $M=100$, and $N=1,024$. In (b) the 
  standard deviation is rather large for the same values of $T$, $t_{rel}$,
  $M$ and $N$ making it difficult to assign a value to $p_s$. However, by
  making $t_{rel}=10^5$ there is a well defined transition at $p_s=0.303$
  as shown in the insert.}
\end{figure}

\begin{figure}
\includegraphics[width=\columnwidth]{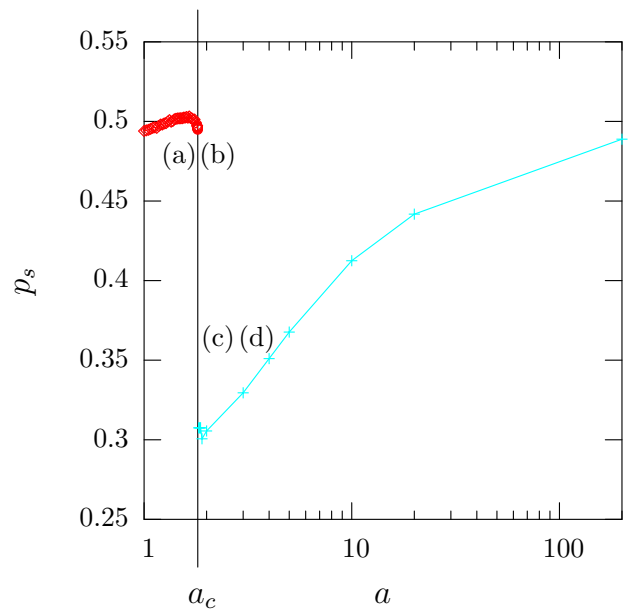}
\caption{\label{fig:ps-a} Synchronization threshold $p_s$ vs. $a$.
 Labels refer to  patterns in Fig.~\ref{fig:sync-cml-00-gr}.}
\end{figure}

Let us now turn to synchronization. In Fig.~\ref{fig:s02}  we show the
dependence of the order parameter $\PROM{u}$ with $p$. For values of
$a$ close to the critical value there is a wide spread in the values
of $u$ as shown in Fig.~\ref{fig:s02} (b) making it difficult to
define  $p_s$. By letting the system to relax longer, we find a well
defined  synchronization threshold as shown in the inset of the
figure. In Fig.~\ref{fig:ps-a} we show the phase diagram of $p_s$ vs.
$a$. Note the large jump of $p_s$ at $a_c$.

\begin{figure}
\begin{tabular}{cc}
 (a) & (b) \\
 \includegraphics[width=0.45\columnwidth]{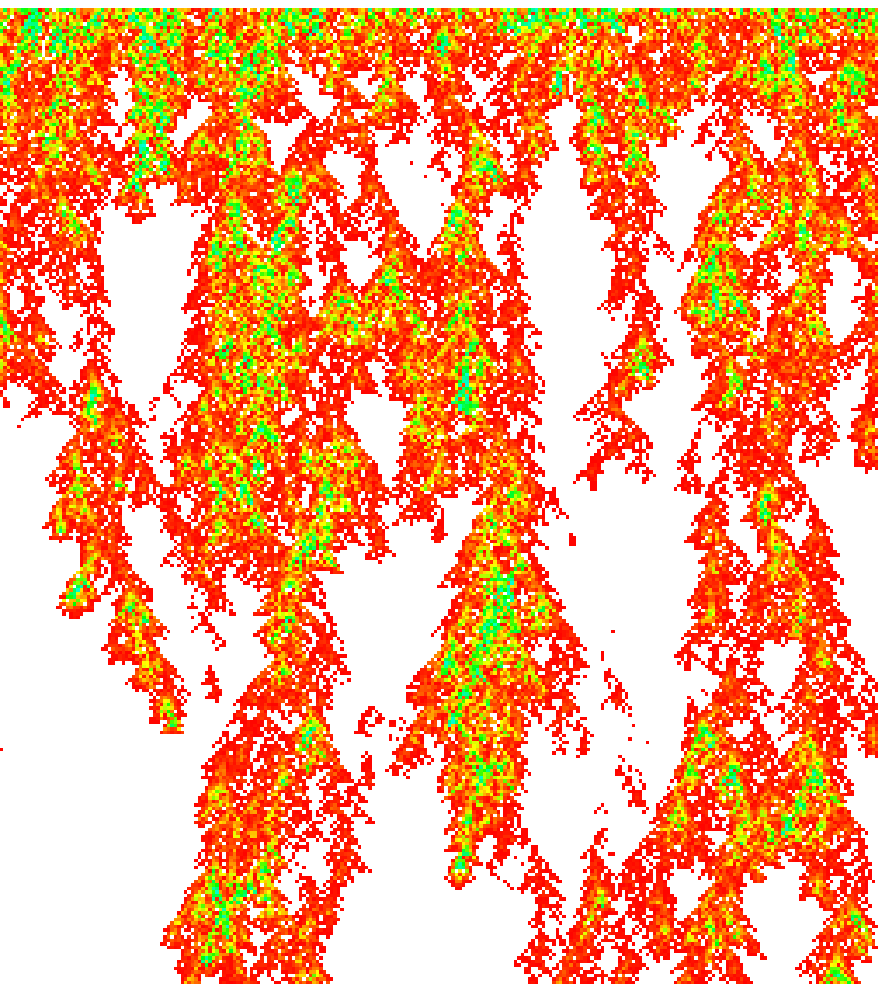} &
 \includegraphics[width=0.45\columnwidth]{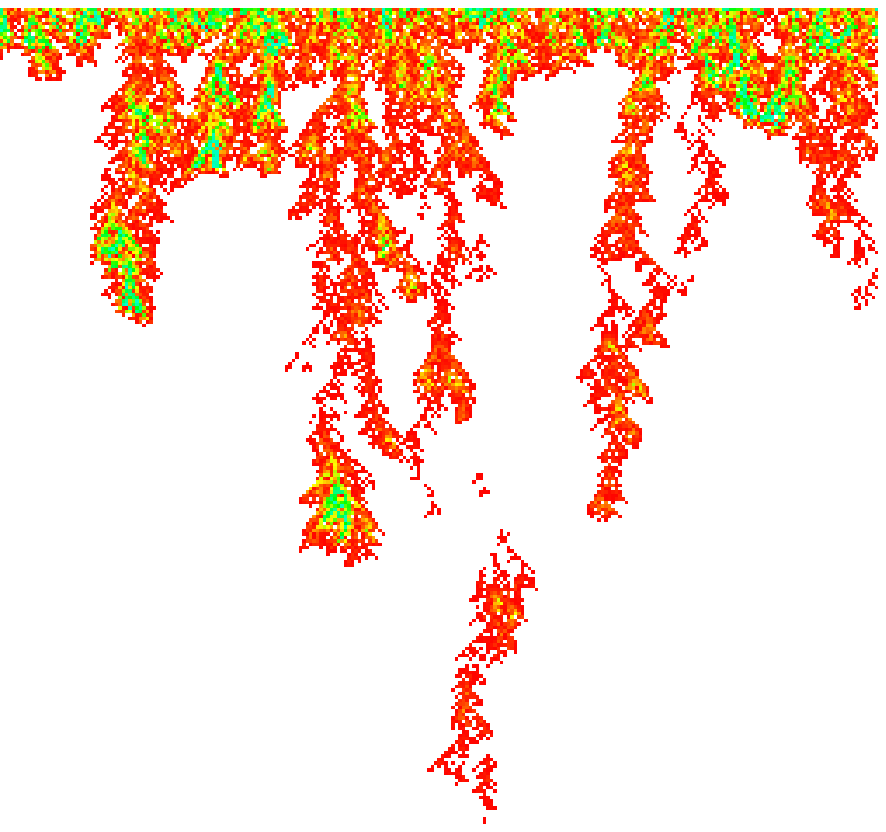}\\
 (c) & (d)\\
 \includegraphics[width=0.5\columnwidth]{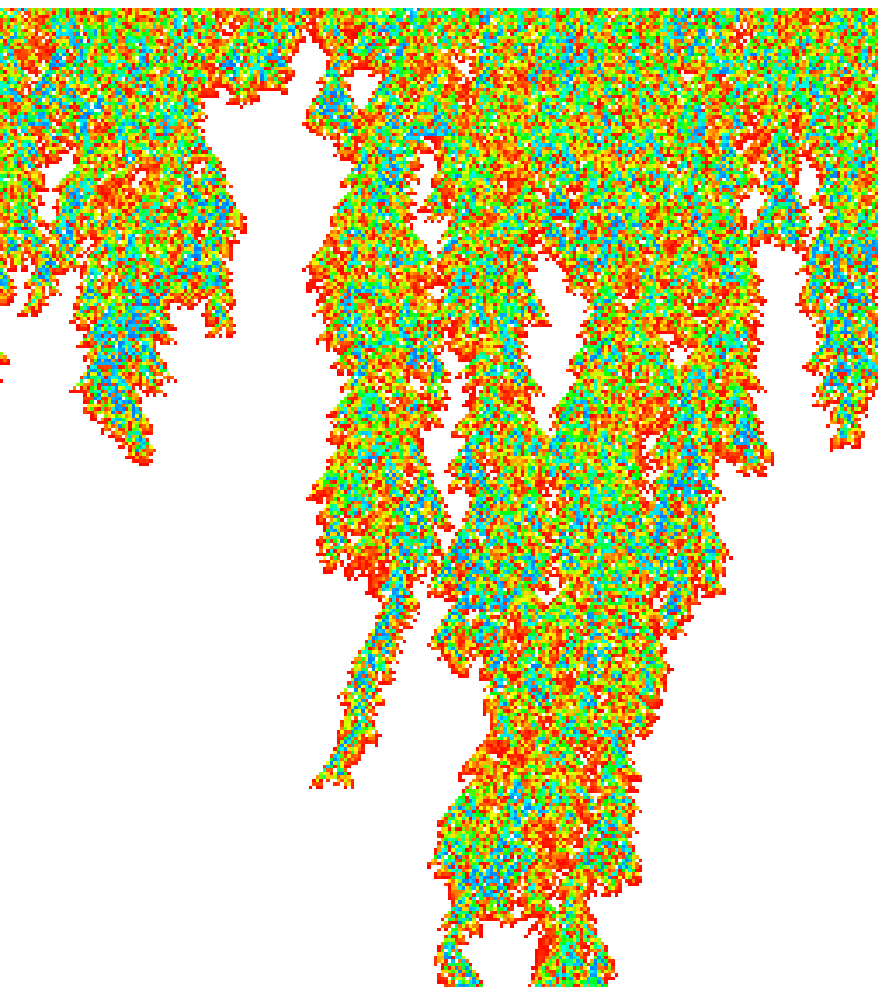}&
 \includegraphics[width=0.5\columnwidth]{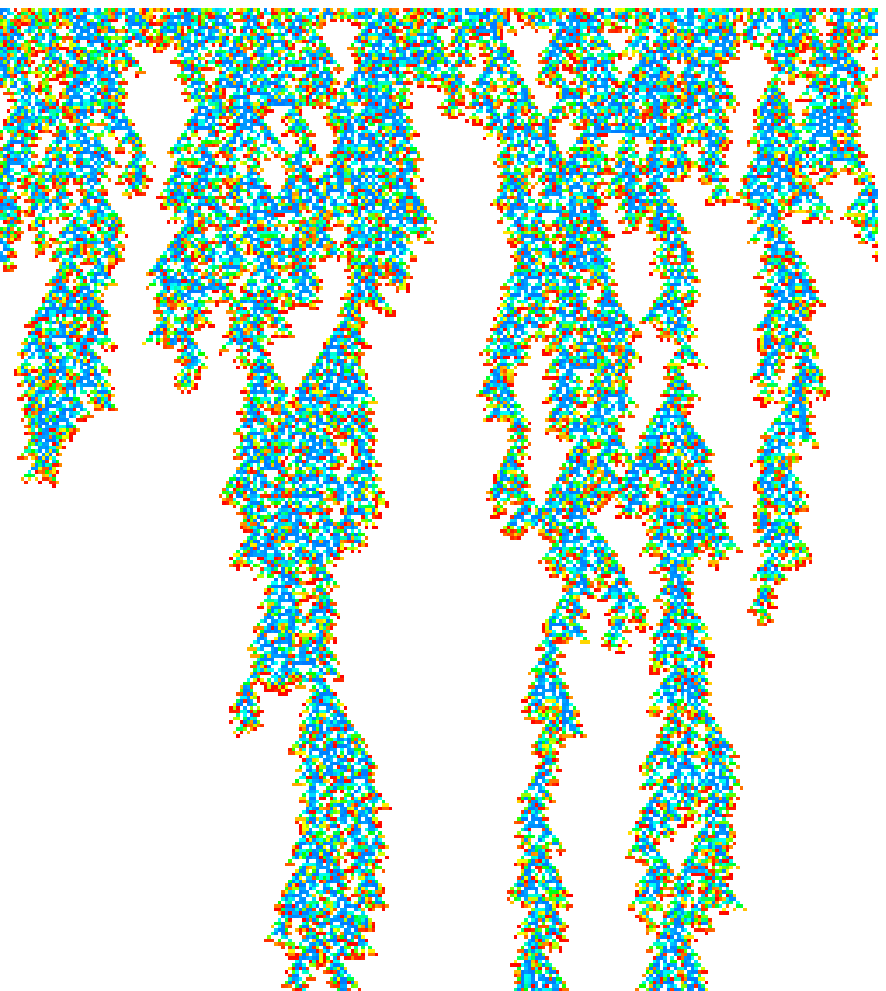}
 \end{tabular}
 \caption{\label{fig:sync-cml-00-gr} (Color Online) Synchronization
  space time  patterns near $p_s$ for different values of $a$. See
  Fig.~\ref{fig:ps-a} for  reference. (a)  $a=1.8<a_c$ ($\lambda >
  0$), $p=0.48<p_s(1.8)$: the difference  field is expanding  but low
  in value, and the synchronized state is not locally  absorbing (b)
  $a=1.9>a_c$ ($\lambda < 0$), $p=0.48\gg p_s(1.9)$: the difference 
  field is contracting, and the synchronized state is absorbing, even
  though  occasionally patches of non-synchronized sites appear.  (c)
  $a=1.9>a_c $,  $p=0.36 \geq p_s(1.9)$: the difference field is
  sustained by the nonlinear term,  the unsynchronized sites form a
  connected cluster. (d) $a=3.5\gg a_c $,  $p=0.36 \geq p_s(3.5)$: the
  linear contribution to the difference field is  (almost) absent, the
  non-synchronized cluster is connected and has more ``holes''  than
  in case (c). $N=256$, $T=256$.}
\end{figure}

For $a<a_c$, $p_s\simeq 0.5$ and for values of $p$ close to this
value, the two  replicas are forced to stay a small distance apart, so
that only the linear (chaotic)  separation is active. The
synchronization transition thus occurs in  agreement with
Eq.~\eqref{eq:ps}, the difference field may be contracting or 
expanding, according with $p$, but the synchronized state is not
absorbing  (panel (a) in Fig.~\ref{fig:sync-cml-00-gr}). By increasing
$a$, the Lyapunov  exponent drops to $-\infty$ (panel (b) in
Fig.~\ref{fig:sync-cml-00-gr}) and we  cross the phase boundary. By 
lowering $p$,  the non-linear mechanism for the diffusion of the
difference field  is activated and sustains the differences between
the replicas. The cluster of  non-synchronized sites is now connected,
like in DP clusters (panel (c) in Fig.~\ref{fig:sync-cml-00-gr}). By 
further increasing $a$, the linear mechanism becomes quickly
contracting, and  the non-synchronized cluster is more reminiscent of
cellular automata ones  (panel (d) in Fig.~\ref{fig:sync-cml-00-gr}). 

\begin{figure}
\begin{center}
 \includegraphics[width=80mm]{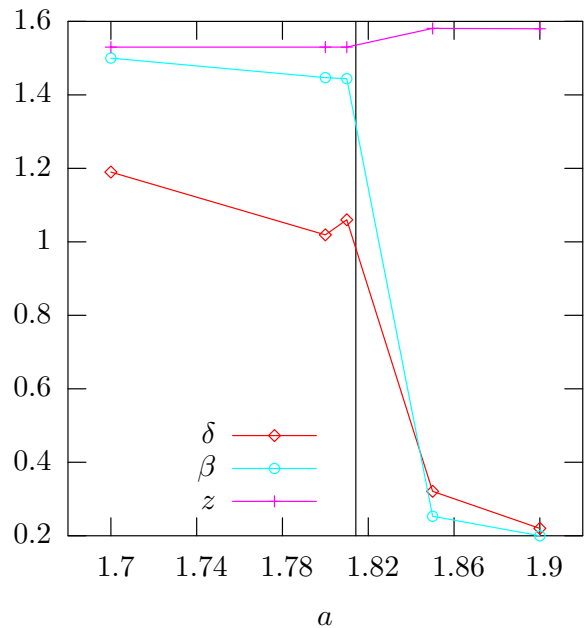}
 \caption{\label{fig:exp-crit} (Color online) The values of the
  critical exponents  $\delta$, $\beta$, and $z$ for different values
  of $a$. The vertical line is  drawn at $a_c=1.8142$.}
\end{center}
\end{figure}

By extensive numerical simulations we found the values of the critical
exponents 
$\delta$, $\beta$ and $z$ defined by
\[
 \PROM{u}\sim (p_s-p)^\beta,\quad \PROM{u^t}\sim t^{-\delta},\quad%
 \PROM{u^t}\sim N^{-\delta z}g(tN^{-z})
\]
with $g$ an unknown function. 
For $a$ below the critical value $a_c$ the synchronization transition
belongs to the bounded Kardar--Parisi--Zhang (BKPZ) universality class
( $\delta =1.10\pm 0.12$, $\beta= 1.50\pm 0.15$, and $z= 1.53\pm
0.07$) and for $a>a_c$ it  belongs to the directed percolation (DP)
universality class ($\delta =0.159464(6)$,  $\beta= 0.276486(6)$ and
$z= 1.580745(1)$ as shown in  Fig.~\ref{fig:exp-crit}. The precision
of the reference values of the  critical exponents reflect the
difficulties in measuring  the critical behavior of systems in the
presence (absence) of absorbing states.  Numerical  experiments
concerning DP are much more precise that those concerning MN, due to
the absorbing character of the void state. In both cases the most
sensitive measure concerns the spreading rate of a defect, but while
in some DP simulations (not in this case)  it is sufficient to
simulate the behavior of the system around the spreading damage (the
evolution of the void state is trivial),  in MN experiments one has to
simulate the whole system. This implies that these DP experiments
immediately reach the asymptotic limit for what concerns the spatial
dimension, while MN experiments suffer from finite-size spatial
effects.  Moreover, DP systems may be discrete, while MN ones are
simulated using floating-point arithmetic. 

\section{Discussion}

Our results suggest that the jump from one universality class to the
other occurs at the value $a_c$ where the jump from chaos to  stable
chaos occurs. In other words, the microscopic stability properties of
the CML are responsible for its macroscopic synchronization
transition. 

Near the synchronization threshold, the difference field behavior is
dominated by the spreading properties of non-synchronized patches,
which is exactly what is generally measured for the determination of
the critical properties of the transition. The evolution of the
difference field, which is in average small at the transition, is
related to the linear stability of the individual map: systems which
are linearly expanding (positive MLE) exhibit a difference field which
is locally small everywhere, with occasional fluctuations. In this
case one has a local competition between the synchronization strength 
and the diverging character of the dynamics. Systems exhibiting stable
chaos, however, are quite different. The difference field cannot be
small, otherwise it would spontaneously vanish due to the negativity
of MLE. Therefore, it has to be large in some patches of small
extension (since at the transition it is small in average), and is
maintained and propagated by the nonlinear part of the dynamics that
amplifies ``instantaneously'' the differences between replicas. This
amplification means that even if two state values in a certain
location of the two replicas are similar, they may become very
different in just one step. This effect can obviously be generated by
the explicit discontinuities in the local evolution
rule~\cite{ahlers02}, but may manifest also in continuous maps, as
shown by our toy model. Actually, our model exhibits this behavior
when the dynamics has reached the ``cellular automaton phase'',
\emph{i.e.}, after the eventual transient chaotic regime. In this CA
phase the effective dynamics is discontinuous, but this behavior is
not easily recognizable just by looking at the local evolution rule.
Notice that DP-like transition can be exhibited also by truly chaotic
systems with discontinuities, like Bernoulli shifts~\cite{ahlers02}. 
Our results are consistent with recent findings using a stochastic
field theory that recovers both MN and DP behavior in a unified way related to the change of stability of the system~\cite{munozpastor}. 

\section{Conclusions}
\label{sec:conclusions}
We have investigated the relation between the type of unpredictability
and   synchronization transition universality classes, using a coupled
map lattice model  where the local dynamics  is given by a continuous
map that can exhibit chaotic or  stable character. This model may
exhibit either chaotic unpredictability or stable  chaos. 

We have analyzed the synchronization properties of two replicas,
coupled  asymmetrically with a parameter $p$. Linear stability
analysis gives an estimate synchronization threshold $p_\ell$ related
to $\lambda$. This is actually the case for smooth chaotic maps. 

We have shown that this \emph{transverse} direction may stay expanding
(\emph{i.e.}, the  system does not synchronize) for $p>p_\ell$. In
this case the synchronized trajectories  are actually absorbing, but
the measure of initial conditions that bring toward them  is
vanishing. 

We expect to observe similar behavior in other, more physical systems,
and in particular in biological ones: the kind of unpredictability
(chaos) observed in biological systems cannot be ascribed to a chaotic
behavior of cells, that in viable conditions are rather stable
unities, but is presumably due to the coupling among them, in a way
that is reminescent of cellular automata ``chaos''.  

Actually, cellular automata have become a widely adopted modeling
tool, due to the simplicity of description of the interaction rules
and the efficient simulating techniques. They have been used to model
a wide variety of systems, from physics to engineering to biology.
However, one may ask where is the transition from the continuous,
microscopic world to the discrete, macroscopic description of cellular
automata. This is exactly the question addressed in this paper, and we
have shown an example in which this transition may be located quite
precisely. 

\section*{Acknowledgments}

Partial economic support from project IN109602 DGAPA--UNAM, the
Coordinaci\'on de la Investigaci\'on Cient\'\i{}fica UNAM, and the
PRIN2003  project {\it Order and chaos in nonlinear extended systems}
funded by MIUR-Italy is acknowledged. We wish to thank the Dept. of
Applied Physics of CINVESTAV-Merida for hospitality under the CONACYT
contract 46709-F. We thank one anonymous referee for his valuable
observations.

\end{document}